%% file: main.tex
\documentclass[sigconf,nonacm]{acmart}
\AtBeginDocument{%
  }

\setcopyright{acmlicensed}
\copyrightyear{2026}
\acmYear{2026}
\acmDOI{XXXXXXX.XXXXXXX}
\acmConference[MSR '26]{Mining Software Repositories 2026}{April 13-14,
  2026}{Rio de Janeiro, Brazil}
\acmISBN{978-1-4503-XXXX-X/2018/06}

\usepackage{pdflscape}
\usepackage{quiver}
\usepackage{enumitem}
\usepackage{placeins}

\newcommand{\rev}[1]{{#1}}
\newcommand{\revv}[1]{{#1}}

\begin{document}

\title{Causal Inference for the Effect of Code Coverage on Bug~Introduction}

\author{Lukas Schulte} \orcid{0000-0001-9336-2075}
\affiliation{%
  \institution{University of Passau}
  \city{Passau}
  \country{Germany}}
\email{lukas.schulte@uni-passau.de}

\author{Gordon Fraser} \orcid{0000-0002-4364-6595}
\affiliation{%
  \institution{University of Passau}
  \city{Passau}
  \country{Germany}}
\email{gordon.fraser@uni-passau.de}

\author{Steffen Herbold} \orcid{0000-0001-9765-2803}
\affiliation{%
  \institution{University of Passau}
  \city{Passau}
  \country{Germany}}
\email{steffen.herbold@uni-passau.de}

\renewcommand{\shortauthors}{Schulte et al.}

\begin{abstract}

  \textbf{Context:} Code coverage is widely used as a software quality assurance measure. However, its effect, and specifically the advisable dose, are disputed in both the research and engineering communities. Prior work reports only correlational associations, leaving results vulnerable to confounding factors. \\
  \textbf{Objective:} We aim to quantify the causal effect of code coverage (\rev{exposure}) on bug introduction (outcome) in the context of mature JavaScript and TypeScript open source projects, addressing both the overall effect and its variance across coverage levels. \\
  \textbf{Method:} We construct a causal directed acyclic graph to identify confounders within the software engineering process, modeling key variables from the source code, issue- and review systems, and continuous integration. Using generalized propensity score adjustment, we will apply doubly robust regression-based causal inference for continuous \rev{exposure} to a novel dataset of bug-introducing and non-bug-introducing changes. We estimate the average treatment effect and dose-response relationship to examine potential non-linear patterns (e.g., thresholds or diminishing returns) within the projects of our dataset.

\end{abstract}




\maketitle

\section{Introduction}

Software quality is a central aspect of the software engineering process, and upholding it in practice serves the purpose of avoiding bugs that make it to production. To this end, test adequacy criteria and code coverage criteria, specifically, are often used as a quality assurance (QA) measure \cite{chenRevisitingRelationshipFault2020}. \rev{Naively, one might assume that well-tested changes perform according to the expectations of the developer and therefore should not introduce bugs}. Within this study, we contribute arguments to this theory that are based on \rev{the causal framework by Judea Pearl \cite{pearl_book_2018} that go beyond correlations. By taking a step on the ladder of causation \cite{pearl_book_2018}, we anticipate that our findings will also aid practitioners in creating decision-making policies.} Concretely, we will consider code coverage as the \rev{exposure} variable in a causal model that describes the software engineering process in mature JavaScript (JS) and TypeScript (TS) open-source projects and quantify its causal effect on bug introduction. We expect our results to shed light on the Average Treatment Effect (ATE)—the effect a one-unit increase in \rev{exposure} has on the probability of bug-introduction—and on the dose response—how the probability of bug-introducing changes across the full range of observed code coverage values.

In observational settings, where controlled experiments are not feasible, causal models can be paired with matching or adjustment strategies to construct methods that approximate controlled experiments \cite{pearlCausalDiagramsEmpirical1995}. In software engineering, and especially in the field of mining software repositories, we operate in such an \rev{observational} setting most of the time. However, prior research has so far mostly neglected this option and relied on the analysis of correlations in data from software repositories as opposed to establishing more robust causal relationships.

Our study complements prior work on bug introduction, like the framework of how bugs are introduced \cite{rodriguez-perezHowBugsAre2020}, as well as QA and defect prediction \cite{kameiLargescaleEmpiricalStudy2013a}. Code coverage, as one form of QA, has been intensively studied as well, for example from the view of testability \cite{ghafariTestabilityFirst2019a}, use in industry \cite{ivankovicCodeCoverageGoogle2019}, automated test generation~\cite{chenRevisitingRelationshipFault2020}, and its correlation with bug introduction \cite{mockusTestCoveragePostverification2009,chenTEvosLargeScaleLongitudinal2023}.

However, causality between code coverage and bug introduction through causal modeling and inference has, to the best of our knowledge, not yet been investigated. Moreover, causal techniques have been underrepresented \rev{and fragmented} in software engineering research as a whole \rev{\cite{siebertApplicationsStatisticalCausal2023}}. With this study, we \rev{investigate this causal link through a causal model} by considering code coverage as a continuous \rev{exposure} for bug introduction, with the higher goal of aiding the discussion of \emph{why} bugs are introduced in open source software. Our expected contributions are the following:

\begin{itemize}
  \item An analysis of the causal effect of code coverage on bug introduction, quantified through the ATE and dose-response functions.
  \item A novel dataset of longitudinal code coverage data, source code metrics, issue and review contents, and bug-introducing changes.
\end{itemize}

To establish consensus on the design of our causal model and analytical approach, we pre-register this study. The pre-registration includes our key assumptions on the relations between variables and the planned methods, preventing post-hoc changes to the study design.

The remainder of this paper is structured as follows. In Section~\ref{sec:related_work}, we provide a short overview of existing related work on causal analysis in software engineering, code coverage, and bug introduction. Next, in Section~\ref{sec:rq}, we introduce our research questions and hypotheses, followed by our research protocol in Section~\ref{sec:protocol} that includes a description of subjects, variables, and the execution plan. Finally, in Sections~\ref{sec:limitations} and \ref{sec:conclusion} we discuss limitations and the conclusion, respectively.

\section{Related Work}
\label{sec:related_work}

Our summary of related work is twofold. First, we describe the state of causal analysis in software engineering research. Second, we give a short overview of code coverage and bug introduction research.

\subsection{Causal Analysis in Software Engineering Research}

Causal modeling, propensity score adjustment, and inference are not yet firmly established as techniques in the field of software engineering. The number of studies that employ these techniques is still small, with the most noteworthy studies being recent ones.

In a comprehensive study into causal analysis of empirical software engineering data, \citeauthor{furiaCausalAnalysisEmpirical2024a}~\cite{furiaCausalAnalysisEmpirical2024a} investigates programmer performance in a coding contest from a Bayesian perspective. In another study, \citeauthor{furiaMitigatingOmittedVariable2025a}~\cite{furiaMitigatingOmittedVariable2025a} provide guidelines to mitigate omitted variable bias in software engineering research through causal analysis. Still in the context of explicit Bayesian analysis, \citeauthor{frattiniApplyingBayesianData2025}~\cite{frattiniApplyingBayesianData2025} refined a study initially conducted by \citeauthor{femmerImpactPassiveVoice2014}~\cite{femmerImpactPassiveVoice2014} with updated causal assumptions and the application of causal modeling and inference.

In contrast to the Bayesian approaches, our study will rely on non-Bayesian methods (i.e., causal modeling, adjustment, and inference)\rev{. We adopt non-Bayesian methods because they allow us to clearly separate the assumptions encoded in our causal model from the data-driven estimation process. This separation enhances both transparency and interpretability (i.e., by making our upfront assumptions explicit).} Here, the field of related work from the software engineering space is even more limited. In fact, we \rev{only found three publications that apply propensity scores: two related publications that use generalized propensity scores for fault localization in numerical software \cite{baiNUMFLLocalizingFaults2015,baiCausalInferenceBased2017} and one that applied propensity score matching on seven projects from industry and open source backgrounds \cite{feyziInforenceEffectiveFault2019}. Aside from that,} causal modeling and inference have recently been used as tools defining test adequacy criteria for computational models by \citeauthor{fosterCausalTestAdequacy2024} \cite{fosterCausalTestAdequacy2024}, as well as metamorphic testing by \citeauthor{clarkMetamorphicTestingCausal2023} \cite{clarkMetamorphicTestingCausal2023}. Yet their contexts differ from ours, since they do not focus on the broad software engineering process.

Further related work applies causal discovery algorithms that support the inference of causal models from data \cite{huPracticalApproachExplaining2023, kazmanCausalModelingDiscovery2017a,yuanImprovedConfoundingEffect2023}. For example, \citeauthor{kazmanCausalModelingDiscovery2017a}~\cite{kazmanCausalModelingDiscovery2017a} applied the PC algorithm to discover a causal graph based on 6 variables: age, \# developers, \# lines of code (LOC), \# design violations, bug churn, and \# bugs. However, our study will not apply causal discovery and instead create the causal model based on domain knowledge.

\subsection{Code Coverage and Bug Introduction}

Our study is most similar to \citeauthor{mockusTestCoveragePostverification2009}~\cite{mockusTestCoveragePostverification2009}, who investigate code coverage and post-verification (i.e., after all relevant tests were executed) defects in a dual case study. They use knowledge gained from interviews to identify variables related to complexity metrics, modification history, and development team setup as potential sources of confounding. However, they do not provide a causal model and only adjust their linear models for the influence of \textit{fan out} (number of methods/functions called added over all methods in a class), \textit{delta} (number of changes to a file in the version control system), and \textit{modification requests} (number of modification requests excluding post-system verification modification requests). From their analyses, they conclude that higher levels of code coverage are associated with better software quality. They also find no evidence of diminishing returns (i.e., when an additional increase in coverage brings a smaller decrease in fault potential). Investigating potential diminishing returns will be part of our study as well.

Other studies investigating code coverage and bug introduction include \citeauthor{ghafariTestabilityFirst2019a}~\cite{ghafariTestabilityFirst2019a}, who investigated testability and its relation to coverage and software quality and found that tested components have fewer bugs, but testability (e.g., negatively affected by violations of the single responsibility principle, unclear correct behavior) strongly influences test quality.
Further, \citeauthor{chenRevisitingRelationshipFault2020}~\cite{chenRevisitingRelationshipFault2020} conducted a review of contradictory analyses regarding the relationship between fault detection, test adequacy (e.g., code coverage), and test set size. Their review highlights that while some studies report a strong correlation between code coverage and fault detection \cite{gopinathCodeCoverageSuite2014}, others, after controlling for test set size, find only a weak or negligible correlation \cite{inozemtsevaCoverageNotStrongly2014a}. Our paper will focus on line-based code coverage, which is a more realistic test adequacy criterion that is also explicitly reported through CI systems and implicitly considered by developers writing tests for their code.

Otherwise, related studies include \citeauthor{ivankovicCodeCoverageGoogle2019}~\cite{ivankovicCodeCoverageGoogle2019}, who researched code coverage usage at Google, and \citeauthor{naminInfluenceSizeCoverage2009}~\cite{naminInfluenceSizeCoverage2009}, who studied test generation and the relationship of test set size on software quality.

To our knowledge, no large-scale longitudinal studies have yet examined organic code coverage (i.e., coverage derived from non-generated tests). A conceptually parallel study, though focused on correlations between test failures and bugs, was conducted by \citeauthor{chenTEvosLargeScaleLongitudinal2023}~\cite{chenTEvosLargeScaleLongitudinal2023}, which analyzed test execution across 12 open-source Java projects. While their dataset includes computed coverage metrics, their compilation success rate was limited to $67.64\%$, and their analysis was capped at $1.000$ commits per project. We aim to base our work on a wider scope by incorporating a larger commit history, more confounding factors, and improving compilation reliability.

In summary, our research is distinct from previous work, as prior studies have primarily focused on identifying correlations between code coverage and bug introduction without establishing causality. While existing literature reports associations, these findings are often confounded by other factors such as code complexity, developer experience, or process-related variables \cite{mockusTestCoveragePostverification2009}. \rev{One exception is the study by \citeauthor{inozemtsevaCoverageNotStrongly2014a} \cite{inozemtsevaCoverageNotStrongly2014a}, who, without explicitly \revv{mentioning} causal methods, measure confounding effects of test suite size on the correlation between coverage and test suite effectiveness.} Our study is designed to \rev{explicitly} estimate the \emph{causal effect} of code coverage on the bug-introducing changes by leveraging a formal causal model, propensity score adjustment, and causal inference. This approach allows us to address confounding and provide more robust evidence regarding the role of code coverage in preventing bugs.



\section{Research Questions and Hypotheses}
\label{sec:rq}

With the aim to quantify the causal effect of code coverage (\rev{exposure}) on bug introduction (outcome), we formulate the following research questions and hypotheses:

\begin{itemize}
  \item[\textbf{RQ1:}] \label{rq:effect} What is the average causal effect of code coverage on bug introduction, and how \rev{strong is the effect across different coverage levels?}
  \item[\textbf{H1:}] Low code coverage has a positive effect on bug introduction, even after accounting for confounding.
  \item[\textbf{H2:}] The probability of bug introduction reduces with increasing code coverage.
  \item[\textbf{H3:}] The reduction of bug introduction probability is non-linear, with a point of diminishing returns after a certain level of coverage is reached.

  \item[\textbf{RQ2:}] \label{rq:confounding} To what extent is the association between coverage and bug introduction attributable to confounding by other bug-introducing factors?
  \item[\textbf{H4:}] The unadjusted association between low code coverage and bug introduction is \rev{different} than the adjusted causal effect, indicating the presence of confounding.

\end{itemize}
We hypothesize that low code coverage has a positive effect on bug introduction, i.e., it increases the probability of bugs (H1). Further, we adapt the hypothesis of \citeauthor{mockusTestCoveragePostverification2009} \cite{mockusTestCoveragePostverification2009} and expect to observe a decrease in probability of bug introduction with rising levels of coverage (H2). However, in line with software "folklore" \cite{mockusTestCoveragePostverification2009}, we expect to see diminishing returns, i.e., progressively smaller decreases in probability of bugs after a certain level of coverage is reached (H3).  We further hypothesize that this effect is overestimated in analyses that do not adjust for confounding (H4).

\section{Research Protocol}
\label{sec:protocol}

The following sections define the subjects, variables, and execution plan of our research protocol.

\subsection{Subjects}

Our subjects are commits from active, mature, open source projects from GitHub that use JS or TS as the primary programming language. The two languages were selected due to their widespread adoption in modern software development and their centralized dependency management through the package managers npm and yarn, which enables the use of the WayPack Machine,\footnote{\url{https://codeberg.org/l-schulte/waypack-machine}} a new tool for temporal reconstruction of npm/yarn dependency installation contexts. \rev{Within those languages we specifically target mature projects since the collection of coverage data relies on the availability of an executable and stable test suite, which may not be the case for newer projects.} We \rev{therefore} pre-filter projects by the following criteria:

\begin{itemize}
  \item Programming language: JavaScript or TypeScript
  \item Number of Commits: min. 10.000
  \item Number of Contributors: min. 10
  \item Number of Stars: min. 100
  \item Last commit: min. 01.08.2025
\end{itemize}

Our preliminary search found $538$ matching projects ($216$ for JS and $322$ for TS) \cite{Dabic:msr2021data}. \rev{Based on an initial assessment, we determined that 20 projects is an achievable sample size that balances manual effort with generalizability to other mature JavaScript/TypeScript projects.} Projects will be randomly sampled from the pre-filtered ones and filtered further based on the following hard inclusion criteria:

\begin{itemize}
  \item The project's development process must be based on feature branches and include code reviews. Code reviews must be hosted on GitHub (i.e., GitHub Pull Requests).
  \item Project commits must be compilable, and code coverage calculation must be executable for at least $90\%$ of commits from the past 5 years.
  \item The code coverage setup must calculate line-based coverage.
  \item Project issues must be hosted on either GitHub, Jira, or BugZilla.
  \item Projects must use GitHub Actions for Continuous Integration~(CI).
\end{itemize}

\rev{}

While dependency resolution will be supported by the WayPack Machine, we still expect significant manual effort to enable coverage data collection. Based on preliminary tests conducted on the Visual Studio Code project\footnote{\url{https://github.com/microsoft/vscode}} we estimate one to three work days \rev{per project} for the creation of \rev{environments in which test suites can be executed and coverage can be extracted. Where possible, coverage will be collected using the tools that are already configured in the project (e.g., Istanbul, Jest, c8). Otherwise, we attempt to configure a tool compatible with the test suite. All coverage data will be collected in the LCOV file format.} The \rev{preliminary} tests were also used to guide our inclusion criteria: for Visual Studio Code, we achieved a success rate (i.e., successful generation of coverage data) of $95\%$ out of a randomly selected $3,000$ commits from the past 5 years.

\begin{figure}
  \centering
  \includegraphics[width=0.45\textwidth]{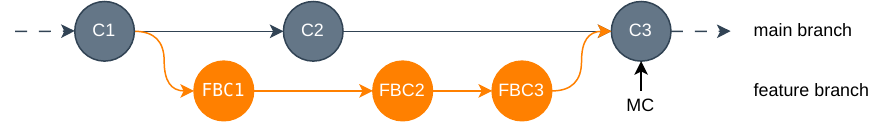}
  \caption{\rev{Visual representation of commits considered in the analysis. Feature-branch commits (FBC) are potential bug-introducing commits, while merge commits (MC) integrate the changes from the FBC into the main branch.}}
  \label{fig:merge}
\end{figure}

We will restrict the commits directly included in the analysis to feature-branch commits (FBC) for which we can identify a merge commit (MC) to the main branch of the project, since source code \rev{change}-related variables are calculated for both the potential bug-introducing commit (i.e., the FBC) and the merge that incorporates all changes from the code review (see \rev{Figure \ref{fig:merge}} and Section \ref{subsec:variables}). \rev{Our subjects therefore always come in pairs: one FBC that is either bug-introducing or non-bug-introducing, and one MC that applies the changes of the feature branch to the main branch.}

\subsection{Variables}
\label{subsec:variables}

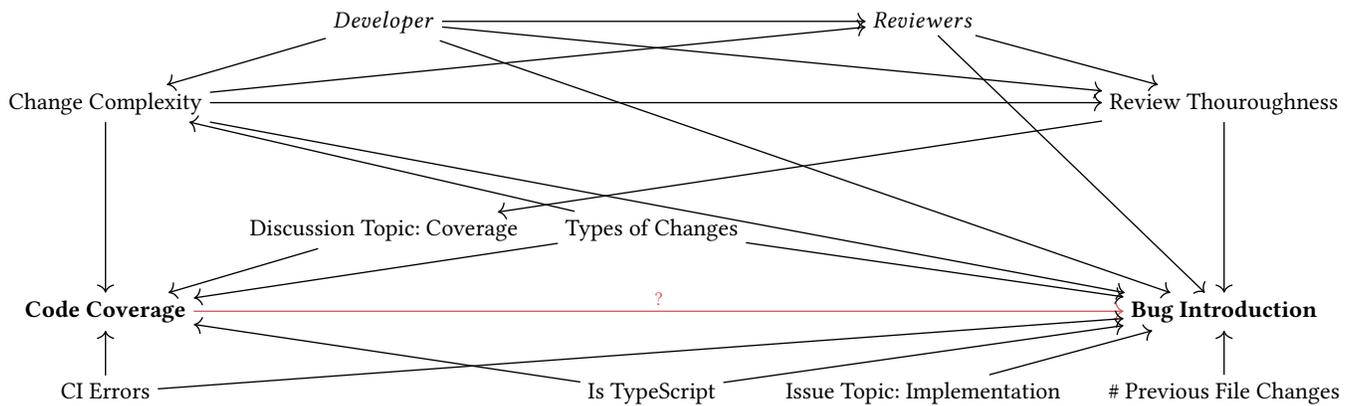
\begin{figure*}
  \input{causal_graph.tex}
  \caption{The proposed causal graph of (latent) variables. Effect between code coverage and bug introduction highlighted in red.}
  \label{fig:causal_graph}
\end{figure*}

The core component of our study is a causal graph (Figure \ref{fig:causal_graph}), which models the software engineering process, focusing on the relationship between code coverage (\rev{exposure}) and bug introduction (outcome). To infer causal relations from the graph's latent variables (e.g., change complexity, review thoroughness), we operationalize them using the observable variables listed in Table \ref{tab:variables}. The selection of variables is based on a study by \citeauthor{schulteExploratoryStudyBugintroducing2023a}~\cite{schulteExploratoryStudyBugintroducing2023a} that provides a broad view on software development practices and their relation to bug introduction. Where possible, we rely on established definitions and automated data collection techniques, i.e., developer-type definitions of \citeauthor{joblinClassifyingDevelopersCore2017a}~\cite{joblinClassifyingDevelopersCore2017a}, the SZZ algorithm \cite{sliwerskiWhenChangesInduce2005}, and refactoring detection with RefDiff 2.0 \cite{silvaRefDiff20MultiLanguage2021}. Further, the collection of variables will be automated where possible, using existing and new tools (e.g., the SmartSHARK platform \cite{trautschAddressingProblemsReplicability2018,trautschSmartSHARKEcosystemSoftware2020}, the WayPack Machine), or custom scripts.


\begin{table*}[]
  \centering
  \caption{List of variables present in the causal graph, including hierarchical abstraction through latent variables.}
  \label{tab:variables}
  \begin{tabular}{lllp{9.35cm}}
    \hline
    \textbf{Variable}                     & \textbf{Type}    & \textbf{Scale} & \textbf{Description}                                                       \\ \hline
    Code Coverage                         & \rev{Exposure}   & Numeric        & Ratio of \rev{lines of} code covered by unit tests in the MC.              \\ \hline
    Bug Introduction                      & Outcome          & Boolean        & Whether the FBC was found to be bug introducing.                           \\ \hline
    Developer                             &                  &                &                                                                            \\
    $\mapsto$ Core Developer              & \rev{Confounder} & Boolean        & Whether the FBC developer was a core developer in the project.             \\
    $\mapsto$ Peripheral Developer        & \rev{Confounder} & Boolean        & Whether the FBC developer was a peripheral developer in the project.       \\
    $\mapsto$ Not Involved                & \rev{Confounder} & Boolean        & Whether the FBC developer was not previously involved in the project.      \\
    $\mapsto$ Code Ownership              & \rev{Confounder} & Numeric        & Ratio of previous commits that the developer made to the FBC source files. \\ \hline
    Types of Changes                      &                  &                &                                                                            \\
    $\mapsto$ Refactoring (FBC)           & Confounder       & Boolean        & Whether the FBC includes a refactoring.                                    \\
    $\mapsto$ Test Change (FBC)           & Confounder       & Boolean        & Whether the FBC includes a test change.                                    \\
    $\mapsto$ Design Change (FBC)         & Confounder       & Boolean        & Whether the FBC includes a design change.                                  \\
    $\mapsto$ Feature Change (FBC)        & Confounder       & Boolean        & Whether the FBC includes a feature change.                                 \\
    $\mapsto$ Refactoring (MC)            & Confounder       & Boolean        & Whether the MC includes a refactoring.                                     \\
    $\mapsto$ Test Change (MC)            & Confounder       & Boolean        & Whether the MC includes a test change.                                     \\
    $\mapsto$ Design Change (MC)          & Confounder       & Boolean        & Whether the MC includes a design change.                                   \\
    $\mapsto$ Feature Change (MC)         & Confounder       & Boolean        & Whether the MC includes a feature change.                                  \\
    \hline
    Change Complexity                     &                  &                &                                                                            \\
    $\mapsto$ Cyclomatic Complexity (FBC) & Confounder       & Numeric        & Cyclomatic Complexity of the changed files in the FBC.                     \\
    $\mapsto$ \# Files Changed (FBC)      & Confounder       & Numeric        & Number of files changed in the FBC.                                        \\
    $\mapsto$ \# Lines Changed (FBC)      & Confounder       & Numeric        & Number of lines changed in the FBC                                         \\
    $\mapsto$ Cyclomatic Complexity (MC)  & Confounder       & Numeric        & Cyclomatic complexity of the changed files in the MC.                      \\
    $\mapsto$ \# Files Changed (MC)       & Confounder       & Numeric        & Number of files changed in the MC.                                         \\
    $\mapsto$ \# Lines Changed (MC)       & Confounder       & Numeric        & Number of lines changed in the MC.                                         \\ \hline
    Reviewers                             &                  &                &                                                                            \\
    $\mapsto$ \# Core Reviewer            & \rev{Mediator}   & Numeric        & The number of core reviewers involved in the code review.                  \\
    $\mapsto$ \# Peripheral Reviewer      & \rev{Mediator}   & Numeric        & The number of peripheral reviewers involved in the code review.            \\
    $\mapsto$ \# First-Time Reviewer      & \rev{Mediator}   & Numeric        & The number of first-time reviewers involved in the code review.            \\
    $\mapsto$ \# Reviewers                & \rev{Mediator}   & Numeric        & The total number of reviewers involved in the code review.                 \\ \hline
    Review Thoroughness                   &                  &                &                                                                            \\
    $\mapsto$ \# Review Comments          & \rev{Mediator}   & Numeric        & The number of review comments in the code review.                          \\
    $\mapsto$ \# Comments                 & \rev{Mediator}   & Numeric        & The number of non-review comments in the code review.                      \\
    $\mapsto$ \# Rounds of Review         & \rev{Mediator}   & Numeric        & The number of rounds of review.                                            \\ \hline
    Discussion Topic: Coverage            & \rev{Mediator}   & Boolean        & Whether coverage was discussed in the issue or review.                     \\ \hline
    \# Previous File Changes              & Predictor        & Numeric        & The number of previous changes to files in the FBC.                        \\ \hline
    Issue Topic: Implementation           & Predictor        & Boolean        & Whether implementation details were discussed in the introducing issue.    \\ \hline
    \# CI Errors                          & Confounder       & Numeric        & The number of CI failures / errors related review.                         \\ \hline
    \rev{Is TypeScript}                   & \rev{Confounder} & \rev{Boolean}  & Whether the project programming language is TS (true) or JS (false).       \\ \hline
  \end{tabular}
\end{table*}

The variables \textit{Test Change}, \textit{Design Change}, and \textit{Feature Change} in Table \ref{tab:variables} are defined as follows:

\begin{itemize}
  \item Test Change: modifications to test files (e.g., \texttt{*.test.js}, \texttt{*.spec.ts}, files in \texttt{/test/} or \texttt{/\_\_tests\_\_/} directories) that are not classified as refactorings.
  \item Design Change: non-refactoring modifications to class definitions, interface declarations, method signatures, or import/export statements in source code files.
  \item Feature Change: all remaining non-refactoring changes to source code files, excluding test changes, design changes, non-source files, and inline documentation.
\end{itemize}

\subsection{Relationships}
\label{subsubsec:relationships}

The causal model we propose in this study is a directed acyclic graph (DAG) that consists of the latent variable presented in Section \ref{subsec:variables}. This section presents arguments for the inclusion of relationships and, where relevant, the exclusion of relationships. Where applicable, we directly cite sources from which we derived the relationships. If no source is given, the relationship is included based on the authors own experience in the field of software engineering.

\vspace{3px}

\noindent\textbf{Code Coverage $\rightarrow$ Bug Introduction:} Describes the causal relation that we are interested in inferring from the comparison dataset.

\noindent\textbf{Types of Changes $\rightarrow$ Bug Introduction:} We hypothesize that the different types of changes (feature, design, refactoring) have an effect on bug introduction that goes beyond the mediators code coverage and change complexity, for example, through unobserved mediators like the availability of definitions for correct behavior~\cite{ghafariTestabilityFirst2019a}.

\noindent\textbf{Types of Changes $\rightarrow$ Code Coverage:} We hypothesize that different types of changes are associated with varying levels of code coverage, as certain types of changes are more likely to prompt developers to adjust tests.

\noindent\textbf{Types of Changes $\rightarrow$ Change Complexity:} Different types of changes are hypothesized to have varying levels of complexity, with feature changes potentially being more complex than refactorings.

\noindent\textbf{Change Complexity $\rightarrow$ Bug Introduction:} More complex changes, characterized by cyclomatic complexity, files modified, or lines changed, are hypothesized to increase the likelihood of bug introduction. These metrics have been frequently used in defect prediction studies~\cite{radjenovicSoftwareFaultPrediction2013}.

\noindent\textbf{Change Complexity $\rightarrow$ Code Coverage:} Complex changes are more likely to be hard to test and therefore receive lower code coverage~\cite{ghafariTestabilityFirst2019a}.

\noindent\textbf{Change Complexity $\rightarrow$ Review Thoroughness:} More complex changes may require more thorough reviews, or even overwhelm reviewers, leading to less effective reviews.

\noindent\textbf{Change Complexity $\rightarrow$ Reviewers:} More complex changes may attract more experienced reviewers.

\noindent\textbf{Developer $\rightarrow$ Bug Introduction:} Developer experience, operationalized through core/peripheral/first-time developer status and code ownership, is expected to influence bug introduction rates. We hypothesize that familiarity with the code leads to fewer bugs.

\noindent\textbf{Developer $\not\rightarrow$ Code Coverage:} We hypothesize that developer experience does not directly influence code coverage. Instead, we expect that it only influences code coverage through change complexity and the review that the change receives.

\noindent\textbf{Developer $\rightarrow$ Change Complexity:} Experienced or peripheral developers may focus on solving issues that are specific and require complex solutions, while first-time developers may start with less complex changes (e.g., good first issues\footnote{\url{https://github.blog/open-source/maintainers/browse-good-first-issues-to-start-contributing-to-open-source}}).

\noindent\textbf{Developer $\rightarrow$ Reviewers:} Developers with less experience may be supervised by more experienced reviewers, while core developers may review each other's code.

\noindent\textbf{Developer $\rightarrow$ Review Thoroughness:} Reviews of changes made by less experienced developers may be more thorough, as reviewers may want to provide more detailed feedback to train the developers.

\noindent\textbf{Reviewers $\rightarrow$ Bug Introduction:} The type and experience of reviewers involved in the code review process may influence the likelihood of bug introduction, as experienced reviewers may be more effective at identifying potential issues.

\noindent\textbf{Reviewers $\rightarrow$ Review Thoroughness:} The type and experience of reviewers involved in the code review process may influence the thoroughness of the review, as experienced reviewers may provide more detailed feedback.

\noindent\textbf{Review Thoroughness $\rightarrow$ Bug Introduction:} More thorough reviews, indicated by more reviewers, review comments, and review rounds, are expected to reduce bug introduction.

\noindent\textbf{Review Thoroughness $\not\rightarrow$ Code Coverage:} We hypothesize that review thoroughness does not directly influence code coverage. Instead, we expect that it only influences code coverage through the discussion of the topic coverage.

\noindent\textbf{Discussion Topic: Coverage $\rightarrow$ Code Coverage:} If coverage is discussed within the issue or during the review, it may lead to changes in code coverage.

\noindent\textbf{CI Errors $\rightarrow$ Bug Introduction:} Continuous integration errors may draw attention to potential issues in the code, thereby reducing the likelihood of bug introduction.

\noindent\textbf{CI Errors $\rightarrow$ Code Coverage:} Continuous integration errors may prompt developers to improve code coverage to address the issues identified.

\noindent\textbf{\# Previous File Changes $\rightarrow$ Bug Introduction:} Files that have been changed more frequently in the past may be more prone to bugs due to their complexity or instability.

\noindent\textbf{Issue Topic: Implementation $\rightarrow$ Bug Introduction:} Issues where implementation details are discussed may be less likely to introduce bugs, as these discussions may lead to better understanding and implementation of the required changes. \rev{On the other hand, discussion may be an indicator complex tasks, which may cause more bugs.}

\noindent\textbf{Is Typescript $\rightarrow$ Code Coverage:} We hypothesize that the programming language (JS or TS) has an effect on code coverage, as language specific aspects like TypeScript's static typing may encourage more comprehensive testing practices, while achieving high levels of coverage is more difficult in JavaScript due to type-related edge-cases.

\noindent\textbf{Is Typescript $\rightarrow$ Bug Introduction:} We hypothesize that the programming language (JS or TS) has an effect on bug introduction rates, as aspects like TypeScript's static typing may help catch certain types of errors at compile time, potentially leading to fewer bugs compared to JavaScript.

\subsection{Execution Plan}

We will start by sampling 20 random projects from the candidate projects and evaluate whether the required data is available. We will then attempt to collect the longitudinal code coverage data using custom scripts and the WayPack Machine. If coverage data collection proves impossible or impractical, we will randomly select another project in its stead.

Once the coverage data has been collected, we will mine the project data using the SmartSHARK platform \cite{trautschAddressingProblemsReplicability2018,trautschSmartSHARKEcosystemSoftware2020}, which already supports mining most of the required software forge data from GitHub and other sources and will be extended where necessary.

Finally, we will assess the inclusion of further projects depending on updated time and effort estimates.

\subsubsection*{\rev{Exposure} Variable (Coverage):}
\label{subsubsec:treatment}
We will consider the \rev{exposure} variable coverage, i.e., the line-based coverage rate for added or modified lines, as a continuous variable between $0\%$ and $100\%$. The coverage rate will be extracted from the collected coverage reports.

\subsubsection*{Outcome Variable (Bug Introduction):}
\label{subsubsec:outcome}
To identify bug-introducing commits, we will execute a modified version of the SZZ algorithm that ignores whitespaces. Since the main reason for mislabeled commits has been identified as mislabeled issues \cite{herzigItsNotBug2013,herboldProblemsSZZFeatures2022c}, we will apply issue type prediction to improve the issue labels \cite{trautschPredictingIssueTypes2023} and through that SZZ algorithm performance. Bug fixes will further be identified through both keywords and links to bug issues \cite{herboldFinegrainedDataSet2022a}. The binary outcome variable will indicate whether a commit introduces a bug.

\subsubsection*{Minimal Adjustment Sets}

To block all backdoor paths between \rev{exposure} and outcome, we use the following minimal adjustment sets: \texttt{\{CI Errors, Change Complexity, \rev{Is TypeScript,} Review Thoroughness, Types of Changes\}} and \texttt{\{CI Errors, Change Complexity, \rev{Is TypeScript,} Review Topic: Coverage, Types of Changes\}}. Controlling for those sets of variables and thereby closing the backdoor paths will allow us to identify the direct causal effect of coverage on bug introduction. \rev{We will compare the results obtained from both adjustment sets to assess the robustness of our findings.}

\subsubsection*{Generalized Propensity Score Calculation}
We will calculate the Generalized Propensity Score (GPS) \cite{hiranoPropensityScoreContinuous2004a} for the continuous \rev{exposure} variable, code coverage, to estimate the probability density of commits receiving a specific level of coverage given the observed covariates from the minimal adjustment set. Because the distribution of the \rev{exposure} variable is not known a priori, we will first assess its distribution \rev{using \revv{visual and statistical} data analysis techniques} and then select an appropriate statistical model (e.g., a linear regression for normally distributed data, a Generalized Linear Model (GLM) with an appropriate link and response family for skewed data, or a Generalized Additive Model (GAM) for more complex or nonlinear distributions) to estimate the GPS.

\subsubsection*{Common Support and Overlap:}
We will assess common support by examining the distribution of the observed coverage values and trimming edges with sparse data. To assess overlap, we will divide coverage into meaningful bins (e.g., quartiles), creating pseudo-\rev{exposure} groups. We will then compare the distribution of propensity scores across these subgroups. Observations in subgroups with poor overlap will be excluded from the further analysis.

\subsubsection*{Average Treatment Effect:}
\hyperref[rq:effect]{\textbf{RQ1}} will be partially addressed by the estimation of the Average Treatment Effect (ATE) \cite{rubinEstimatingCausalEffects1974} of code coverage on bug introduction. The ATE represents the average causal effect of changing code coverage on the probability of bug introduction across the entire population of commits, quantifying the expected difference in the probability of bug introduction between commits with higher versus lower coverage levels. To this end we will transform the GPS into Inverse Probability Weights (IPW) \cite{naimiConstructingInverseProbability2014a} and fit a double robust weighted regression model based on the \rev{exposure}, the minimal adjustment set, and predictors of the outcome, weighted by the IPW. The appropriate model family (e.g., weighted logistic regression, or GAM) will be chosen based on exploratory data analysis. The inclusion of the minimal adjustment set ensures double robustness by reducing reliance on correct model specification, i.e., the estimator remains consistent even if either the propensity score model or the outcome model is misspecified~\cite{huntington-kleinEffectIntroductionResearch2022a}.

\subsubsection*{Causal Dose-Response Curve:} \label{subsubsec:causal_effect_estimation} To further address \hyperref[rq:effect]{\textbf{RQ1}}, we will estimate the causal dose-response curve \cite{hiranoPropensityScoreContinuous2004a} using our double robust regression model, which will allow us to examine how the probability of bug introducing changes across the full range of code coverage values. The dose-response curve provides a pointwise estimate of the expected probability of bug introduction at each possible coverage value, accounting for confounding by weighting each observation using the GPS/IPW.

We will generate dose-response estimates across the spectrum of observed coverage, and visualize them to illustrate the functional relationship between coverage and bug risk.

\subsubsection*{Effect of Confounding:} To address \hyperref[rq:confounding]{\textbf{RQ2}}, we will compare our results to a naive (unadjusted) regression to assess the bias introduced by confounding. Specifically, we will determine the unadjusted bias with a naive logistic regression: $ logit(P(Bug=1)) = \beta_0 + \beta_1 Coverage $ where $ \beta_{1} $ is the unadjusted effect of coverage. We will compare the unadjusted effect to the ATE and also generate dose-response estimates of the naive regression to overlay with the causal dose-response curve.


\section{Limitations}
\label{sec:limitations}

Our study has four key limitations. First, while we improve bug-introducing commit identification using issue type prediction, the SZZ algorithm remains unreliable with non-negligible error rates. Second, our causal graph assumes specific confounder relationships, which are subjective. We are addressing this via pre-registration of this study. Third, our future findings will be specific to mature JS/TS projects and may not generalize to other contexts. Finally, within the scope of this study, we do not apply methods that control for unobserved confounding and instead leave establishing these techniques in the field of software engineering to future work.

\section{Conclusion}
\label{sec:conclusion}

We will conduct a study that aims to quantify the causal effect of code coverage on bug introduction using a new dataset and techniques for causal modeling and inference not yet established in the field of software engineering. While previous studies have investigated links between code coverage and bug introduction from a correlational perspective, our goal is to check if their findings hold in a causal context and to provide new findings on the dose-response relation between the two variables. \rev{We anticipate that our findings will provide valuable arguments for discussions about code coverage on quality assurance and influence decision-making policies in practice.}

\subsection{Future Work}

While this study focuses on the ATE and the dose-response, future work may explore heterogeneity in the \rev{exposure} effect by subgroup (e.g., code complexity) and different \rev{exposures} all together. Further, establishing techniques for quantifying unobserved confounding will be an important future contribution.

\section{Publication of Data}
\label{sec:data}
All materials that we collect will be \rev{pseudonymized\revv{.} The key for the pseudonyms will only be available for the researchers and the data will be anonymized prior to publication in a long-term archive.}


\bibliographystyle{ACM-Reference-Format}
\bibliography{main}

\appendix

\end{document}

%% file: causal_graph.tex
\[\begin{tikzcd}[cramped,column sep=1.4em,row sep=1.9em]
        & Developer && Reviewers \\
        {\text{Change Complexity}} &&&& {\text{Review Thouroughness}} \\
        \\
        & {\text{Discussion Topic: Coverage}} & {\text{Types of Changes}} \\
        {\textbf{Code Coverage}} &&&& {\textbf{Bug Introduction}} \\
        {\text{CI Errors}} && {\text{Is TypeScript}} & {\text{Issue Topic: Implementation}} & {\text{\# Previous File Changes}}
        \arrow[from=1-2, to=1-4]
        \arrow[from=1-2, to=2-1]
        \arrow[from=1-2, to=2-5]
        \arrow[from=1-2, to=5-5]
        \arrow[from=1-4, to=2-5]
        \arrow[from=1-4, to=5-5]
        \arrow[from=2-1, to=1-4]
        \arrow[from=2-1, to=2-5]
        \arrow[from=2-1, to=5-1]
        \arrow[from=2-1, to=5-5]
        \arrow[from=2-5, to=4-2]
        \arrow[from=2-5, to=5-5]
        \arrow[from=4-2, to=5-1]
        \arrow[from=4-3, to=2-1]
        \arrow[from=4-3, to=5-1]
        \arrow[from=4-3, to=5-5]
        \arrow["{{?}}", color={rgb,255:red,214;green,92;blue,92}, from=5-1, to=5-5]
        \arrow[from=6-1, to=5-1]
        \arrow[from=6-1, to=5-5]
        \arrow[from=6-3, to=5-1]
        \arrow[from=6-3, to=5-5]
        \arrow[from=6-4, to=5-5]
        \arrow[from=6-5, to=5-5]
    \end{tikzcd}\]